\documentclass[%
 reprint,
 amsmath,amssymb,
 aps,
prb,
showkeys
]{revtex4-2}

\usepackage{graphicx}
\usepackage{dcolumn}
\usepackage{bm}
\usepackage{multirow}

\usepackage{graphicx}
\usepackage{siunitx}  
\DeclareSIUnit\angstrom{\protect \text {\AA}}
\usepackage[version=4]{mhchem}
\usepackage{breakurl}
\usepackage{libertine}
\usepackage[libertine]{newtxmath}
\usepackage{float}
\usepackage{booktabs}
\usepackage[
colorlinks=true,      
linkcolor=black,       
citecolor=black,       
filecolor=black,       
urlcolor=black,       
pdfpagemode=UseNone  
]{hyperref}
\usepackage{xcolor}

\begin{document}


\title{Unveiling the thermal transport properties of Biphenylene nanotubes:\\ A molecular dynamics study}

\author{Jhionathan de Lima}
\author{Cristiano Francisco Woellner}
\affiliation{Department of Physics, Federal University of Paraná UFPR, Curitiba, PR, Brazil.}


\begin{abstract}
Biphenylene nanotubes (BPNNTs) represent a novel class of carbon-based nanomaterials, constructed by rolling a biphenylene network (BPN) monolayer into a one-dimensional tubular structure. In this study, the thermal transport properties of BPNNTs are investigated using reverse non-equilibrium molecular dynamics simulations. At room temperature, the lattice thermal conductivity of armchair and zigzag BPNNTs is found to be approximately \SI{100}{\watt\per\meter\per\kelvin} and \SI{90}{\watt\per\meter\per\kelvin}, respectively. These values are at least one order of magnitude lower than those reported for single-walled carbon nanotubes (SWCNTs). This significant reduction is attributed to the unique atomic arrangement of BPNNTs, which leads to a substantially lower phonon group velocity. Furthermore, the effects of nanotube length, diameter, and temperature on thermal transport are systematically analyzed. To elucidate the mechanisms underlying the geometry- and temperature-dependent thermal behavior, a comprehensive analysis of phonon dispersion relations, vibrational density of states, and phonon group velocities is conducted. This study offers valuable insight into the thermal transport properties of BPNNTs, with implications for thermal management and energy-related applications.
\end{abstract}

\maketitle


\section{Introduction}
\label{sec:introduction}
The increasing demand for miniaturized electronic devices and components, driven by constraints on urban space, limited natural resources, and the imperative for clean energy and sustainable development, has significantly accelerated the advancement of nanotechnology~\cite{Malik2023}. Consequently, the study of thermal transport in nanomaterials has attracted substantial attention due to its fundamental role in thermal management and energy-related applications~\cite{Qiu2020, HabeebDolapoSalaudeen2024}.

At the nanoscale, heat conduction, primarily governed by phonons, deviates significantly from macroscopic behavior due to factors such as interfaces, quantum confinement, and structural topology, all of which strongly influence phonon dynamics and, consequently, thermal conductivity~\cite{Xie2023, Cahill2001}. These unique thermal transport characteristics make nanomaterials highly promising for next-generation technologies, including high-performance nanoelectronic devices that require efficient thermal dissipation and thermoelectric materials designed for enhanced energy conversion efficiency~\cite{Pop2010, Shi2012}. A comprehensive understanding of these thermal properties is imperative for unlocking the full potential of nanomaterials in future technological advancements.

Among nanomaterials, two-dimensional (2D) carbon-based structures hold particular significance. Since the experimental isolation of graphene in 2004~\cite{Novoselov2004}, its remarkable thermal conductivity~\cite{Balandin2008, Ghosh2008} and other unique physicochemical properties~\cite{Geim2007the} have driven extensive research into alternative carbon allotropes. Although the majority of newly proposed structures in the literature remain theoretical, recent advancements in experimental techniques have enabled the successful synthesis of several novel structures, including the biphenylene network (BPN)~\cite{Fan2021}, $\gamma$-graphyne~\cite{Li2018, Yang2019, Barua2022}, $\gamma$-graphdiyne~\cite{Li2010}, and the fullerene network~\cite{Hou2022}.

BPN is a novel 2D carbon material characterized by a lattice structure composed of 4-, 6-, and 8-membered carbon rings, with $\mathrm{sp^2}$ hybridization. Extensive theoretical investigations have been conducted to elucidate its electronic~\cite{Bafekry2021, Hou2023, Luo2021}, mechanical~\cite{Luo2021, HamedMashhadzadeh2022, Pereira2022, Mortazavi2022}, and thermal properties~\cite{HamedMashhadzadeh2022, Mortazavi2022, Veeravenkata2021, Zhang2021, Ying2022, Yang2023}. For instance, Mashhadzadeh \textit{et al.}~\cite{HamedMashhadzadeh2022} employed non-equilibrium molecular dynamics (NEMD) simulations with the optimized Tersoff potential to investigate the thermal conductivity of BPN. They reported intrinsic values of \SI{146.41}{\watt\per\meter\per\kelvin} and \SI{133.51}{\watt\per\meter\per\kelvin} for zigzag and armchair monolayers, respectively. Ying \textit{et al.}~\cite{Ying2022} further analyzed the thermal transport properties of BPN in comparison to graphene using the optimized Tersoff force field and three different molecular dynamics-based methods, including homogeneous non-equilibrium molecular dynamics (HNEMD), equilibrium molecular dynamics (EMD), and NEMD. Their study revealed that the thermal conductivity of BPN is approximately $1/13$ that of graphene. This significant reduction was attributed to decreased structural symmetry, which decreases phonon group velocity and shortens the phonon mean free path. Similarly, Harish \textit{et al.}~\cite{Veeravenkata2021} employed DFT calculations to investigate the thermal conductivity of BPN in comparison to graphene. Their study demonstrated that the reduced crystal symmetry of BPN enhances anharmonicity, resulting in a thermal conductivity more than an order of magnitude lower than that of graphene.

Beyond investigations on the 2D BPN structure, research has also been conducted on its one-dimensional (1D) counterpart, biphenylene nanotubes (BPNNTs), focusing on their electronic~\cite{Hudspeth2010} and mechanical~\cite{Armando2024} properties, as well as their potential applications in CO$_2$ gas sensing~\cite{Esfandiarpour2022} and sodium-ion batteries~\cite{Vafaee2023}. However, the existing literature lacks comprehensive studies on the thermal properties of BPNNTs, particularly regarding the influence of geometric parameters on heat transport. Given the uniquely arranged non-benzenoid structure of BPN, which differs from the pristine honeycomb lattice of graphene, BPNNTs are expected to exhibit distinct thermal behavior compared to conventional single-walled carbon nanotubes (SWCNTs). The present study aims to address this gap by systematically investigating the thermal transport properties of BPNNTs, providing insights into their potential applications in nanoscale thermal management.

In this study, the thermal properties of BPNNTs are investigated using reverse non-equilibrium molecular dynamics simulations (RNEMD). The effects of nanotube length, diameter and temperature on thermal transport are examined. Spectral analysis of the phononic properties of BPNNTs, including phonon dispersion, group velocity, and vibrational density of states (VDOS), is conducted to elucidate the underlying mechanisms governing their thermal conductivity. 

\section{Simulation details}\label{sec:methodology}
\subsection{Biphenylene-Based Nanotube Modeling}
\begin{figure}[h!]
	\centering
	\includegraphics[width=1.0\linewidth]{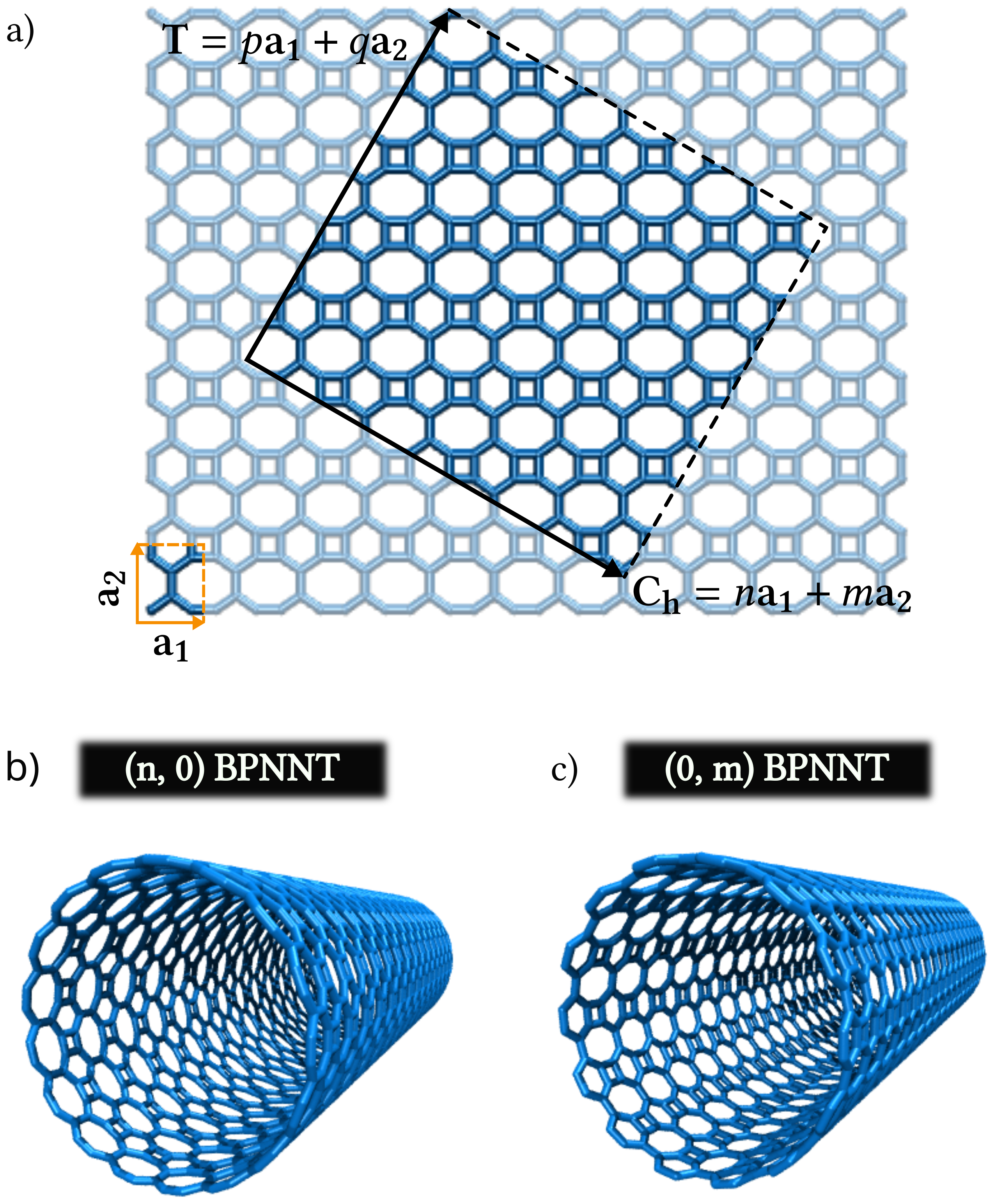}
    \caption{Schematic representation of (a) the BPN monolayer and its unit cell (orange rectangle). The BPNNT naming scheme \((n,m)\) is defined by the chiral vector \(\mathbf{C_h}\) on an infinite BPN sheet, determining the rolling direction to form the nanotube. The vector \(\mathbf{T}\) denotes the tube axis, while \(\mathbf{a_1}\) and \(\mathbf{a_2}\) are the unit vectors in real space. Perspective views of (b) zigzag BPNNT with chiral index $(n,0)$ and (c) armchair BPNNT with chiral index $(0,m)$ are also shown.}
	\label{fig:nanotubes}
\end{figure}

The unit cell of BPN consists of six carbon atoms and exhibits a rectangular shape, with optimized lattice constants of $a_1=\SI{3.76}{\angstrom}$ and $a_2=\SI{4.52}{\angstrom}$~\cite{Fan2021, Veeravenkata2021}, as depicted in \autoref{fig:nanotubes}a. The construction of a BPNNT follows a procedure analogous to that used for SWCNTs. A schematic representation of the vectors involved in this process is shown in \autoref{fig:nanotubes}a. The unit cell is first periodically replicated in the planar directions. The rolling direction of the monolayer is defined by the chiral vector
$\mathbf{C_h}=n\mathbf{a_1}+m\mathbf{a_2}$, where $\mathbf{a_1}$ and $\mathbf{a_2}$ are the orthogonal lattice vectors of the system, and $n$ and $m$ are integers. The chiral vector determines the nanotube circumference, with the diameter given by $d(n,m) = |\mathbf{C_h}|/\pi$. The BPNNT structure is further defined by the translational vector $\mathbf{T}=p\mathbf{a_1}+q\mathbf{a_2}$, where $p$ and $q$ are integers. This vector is the shortest possible vector orthogonal to $\mathbf{C_h}$,  defining the nanotube length as $L=|\mathbf{T}|$. A necessary condition for $p$ and $q$ is given by $\mathbf{C_h} \cdot \mathbf{T} = 0$, which simplifies to $(n \mathbf{a_1} + m\mathbf{a_2}) \cdot (p \mathbf{a_1} + q \mathbf{a_2}) = 0$. Since $\mathbf{a_1} \perp \mathbf{a_2}$, the relationship between these parameters is expressed as $ n p |{\mathbf{a_1}|^2} + m q |\mathbf{a_2}|^2 = 0$. This condition ensures that $m$ and $n$ cannot be simultaneously zero. When $ m = 0 $, it follows that $ p = 0 $, allowing $ q $ to take any integer value. In this case, the chiral vector simplifies to $\mathbf{C_h} = n\mathbf{a_1}$, resulting in a nanotube with a zigzag edge, a diameter of $d{(n,0)} = n|\mathbf{a_1}|/\pi$ and a length of $ L = q|\mathbf{a_2}| $. Similarly, for $ n = 0 $, the nanotube has an armchair edge, a diameter of $d{(0,m)} = m|\mathbf{a_2}|/\pi$ and a length of $ L = p|\mathbf{a_1}| $.

Chiral nanotubes can be generated when the ratio $p/q \approx - (m/n)(|\mathbf{a_2}|/\mathbf{|a_1|})$ holds, which is only possible for considerably high integers $p, q, m, n$. This leads to nanotubes with large diameters. In our approach, with periodic boundary conditions, such configurations occur in systems with a large number of atoms. Given our computational limitations, this study focuses exclusively on armchair and zigzag BPNNTs.

The BPNNT structures were generated using a Python script based on the Atomic Simulation Environment (ASE) library~\cite{HjorthLarsen2017the}. The nanotubes were constructed by rolling the monolayer along the chiral vectors $(n,0)$ and $(0, m)$, resulting in zigzag (\autoref{fig:nanotubes}b) and armchair (\autoref{fig:nanotubes}c) BPNNTs, respectively.

\subsection{Molecular Dynamics Simulations}
To investigate the thermal properties of BPNNTs, molecular dynamics (MD) simulations were performed using the Large-scale Atomic Molecular Massively Parallel Simulator (LAMMPS) package~\cite{Plimpton1995}. The atomic interactions in BPNNTs were described using the Tersoff potential~\cite{Tersoff1988} with a parameter set optimized by Lindsay and Broido~\cite{Lindsay2010}. This potential has been widely utilized to study the thermal properties of carbon-based nanomaterials, including graphene~\cite{Zou2016}, SWCNTs~\cite{Cao2012}, and BPN~\cite{HamedMashhadzadeh2022, Ying2022}. In comparison to the original Tersoff~\cite{Tersoff1988} and the Brenner~\cite{Stuart2000, Brenner2002} potentials, the optimized Tersoff potential provides a more accurate representation of the phonon dispersion in BPN~\cite{Ying2022}. The equations of motion were integrated using the velocity Verlet algorithm~\cite {Martys1999} with a time step of \SI{0.1}{\femto\second}. Periodic boundary conditions were applied along the axial direction of the nanotubes, while a vacuum of \SI{100}{\angstrom} was introduced in the transverse directions to prevent spurious interactions. The initial equilibrium configurations were obtained through energy minimization via the conjugate gradient method. The systems were then thermalized at a finite temperature using a Nosé–Hoover thermostat~\cite{Nos1984, Hoover1985} for \SI{100}{\pico\second}. Subsequently, they were relaxed in an NPT ensemble under stress-free conditions for $1\times 10^6$ time steps. After equilibration, the thermostat was deactivated, and the systems were allowed to evolve under the microcanonical ensemble conditions (NVE) for further thermal conductivity calculations. 

\begin{figure}[h!]
	\centering
	\includegraphics[width=1.0\linewidth]{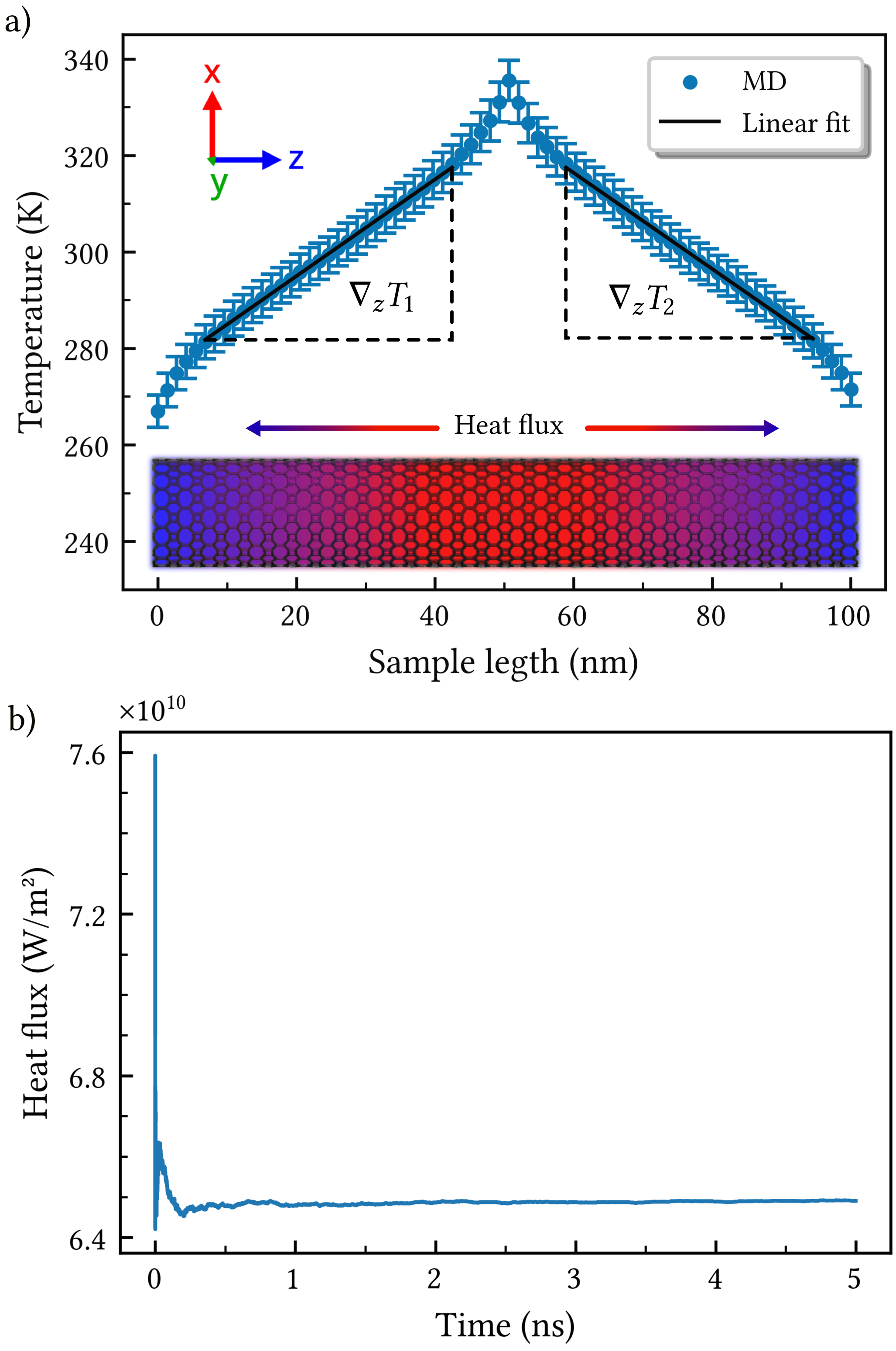}
	\caption{(a) Setup for the RNEMD method and typical temperature profile for an armchair (0,14)-BPNNT with a nominal length of \SI{100}{\nano\meter}. The heat flux is generated along the tube direction by exchanging kinetic energy between slow-moving particles in the hot (red) region and fast-moving particles in the cold (blue) region. The data points, with their respective error bars, represent the average temperature of each slab into which the system is divided. (b) Corresponding heat flux required to maintain the temperature gradient shown in (a).}
	\label{fig:setup}
\end{figure}

The thermal conductivity of BPNNTs is determined using the RNEMD method, based on the M\"{u}ller–Plathe approach~\cite{MllerPlathe1997}. This method involves imposing a heat flux on the system and measuring the resulting temperature gradient. \autoref{fig:setup}a shows the setup of RNEMD simulations performed in this study. The system is divided into $n$ slabs along its length, with two of them assigned as ``hot'' and ``cold'' regions. In our setup, the first slab was assigned as the cold region, while the middle slab served as the hot region. Due to periodic boundary conditions, slabs 0 and $n$ are identical. Each slab contained approximately 300 atoms.

The heat flux is induced by exchanging the kinetic energy of slow-moving particles in the hot region with fast-moving particles in the cold region, thereby generating a temperature gradient. Kinetic energy swaps were performed every 1000 simulation steps over a total of $50\times 10^6$ steps. The heat flux $J$, defined as the energy transferred per unit time $\Delta t$ across a surface perpendicular to the heat flux direction, is given by \cite{MllerPlathe1997}:
\begin{equation}
    J(t)=\frac{1}{2 A \Delta t} \sum_\mathrm{swaps} \frac{m}{2} \left( v_\mathrm{hot}^2 - v_\mathrm{cold}^2  \right),
    \label{eq:heatflux}
\end{equation}
where $m$ is the mass of the carbon atoms, $v_\mathrm{hot}$ and $v_\mathrm{cold}$ are the velocities of the faster-moving atoms in the cold slab and the slower-moving atoms in the hot slab, respectively. The parameter $A$ represents the cross-sectional area of the nanotube. Following the standard convention, the nanotube is treated as a hollow cylinder with a diameter $d$ and a wall thickness of $h=\SI{3.35}{\angstrom},$ yielding a cross-sectional area of $\pi dh$. For a sheet, the cross-sectional area is given by $wh$ where $w$ is the sheet width. The factor of 2 in the denominator of \autoref{eq:heatflux} accounts for the periodicity of the system. The definition of heat flux given in \autoref{eq:heatflux} avoids virial decomposition of the stress tensor, a well‑known limitation of LAMMPS’s analytical heat flux calculation for many‑body potentials \cite{Fan2015, Boone2019}.

At each simulation step, the instantaneous temperature $T_i$ of the $i$-th slab is determined using the equipartition theorem~\cite{MllerPlathe1997}:
\begin{equation}
    T_i=\frac{1}{3 N_i k_B} \sum_{j=i}^{N_i} \frac{p_j^2}{m_j},
    \label{eq:temperature}
\end{equation}
where $N_i$ is the number of atoms in the $i$-th slab, $k_B$ is the Boltzmann constant, $m_j$ is the mass of atom $j$, and $p_j$ is its linear momentum. The temperature profile of the system is then obtained from the average temperature in each slab. \autoref{fig:setup}a illustrates the temperature profile for a sample with a nominal length of \SI{100}{\nano\meter}. The profile exhibits a well-defined symmetry with respect to the center of the sample. Nonlinearity is observed near the hot and cold regions, which can be attributed to finite size effect~\cite{Schelling2002c}. \autoref{fig:setup}b displays the corresponding heat flux required to maintain the temperature gradient shown in \autoref{fig:setup}a. It is evident that a stationary state is reached after \SI{1}{\nano\second}. Even for the largest systems examined, simulation durations of up to \SI{5}{\nano\second} were sufficient to reach a stationary state.

Once a linear temperature gradient and a stationary heat flux are achieved, the thermal conductivity $\kappa$ for a sample of size $L$ is determined using Fourier law of heat conduction:
\begin{equation}
    \kappa(L)=-\frac{\langle J \rangle}{\left \langle \nabla_z T \right\rangle},
    \label{eq:fourier}
\end{equation}
where $\left \langle \nabla_z T \right\rangle$ is the arithmetic mean of the temperature gradient considering both directions of heat flux, as shown in \autoref{fig:setup}a. The brackets $\left \langle \cdot \right\rangle$ denote time-averaged quantities.

\subsection{Spectral analyses}
To gain a better understanding into thermal transport in the BPNNTs considered in this study, their phonon properties and VDOS were analyzed. In MD simulations, the VDOS for a system with $N$ atoms is computed by performing a Fourier transform on the atomic velocity autocorrelation function (VACF)~\cite{Dickey1969}:
\begin{equation}
    \mathrm{VDOS}(\nu)=\int_0^\infty \bigg \langle\dfrac{1}{N}\sum_{j=1}^N\frac{\textbf{v}_j(t)\cdot \textbf{v}_j(0)}{\textbf{v}_j(0)\cdot \textbf{v}_j(0)} \bigg\rangle e^{-2\pi i\nu t}dt,
    \label{eq:vdos}
\end{equation}
where $\mathbf{v}_j$ is the velocity of the $j$-th atom, $\nu$ is the frequency, and $\langle \mathbf{v}_j(t)\cdot \mathbf{v}_j(0) \rangle$ is the VACF, which is normalized such that $\text{VACF}(t=0)=1$. In our simulations, the VDOS was obtained by post-processing \SI{10}{\pico\second} trajectories, where atomic velocities were recorded every $\SI{5}{\femto\second}$. 

Phonon dispersion relations were obtained using the finite displacement method as implemented in the PHONOPY package~\cite{Togo2015}, with input data derived from LAMMPS simulations. Initially, structural relaxation was performed to minimize residual atomic forces. The energy minimization was considered converged when the total energy change was negligible and the maximum force on any atom was below \SI{e{-8}}{\electronvolt\per\angstrom}. Subsequently, supercells incorporating small atomic displacements were generated. For each supercell configuration, interatomic forces were computed. The force constants obtained were then utilized to construct and diagonalize the dynamical matrix, yielding phonon dispersion relations and phonon group velocities.

\section{Results and discussion}
\subsection{Phonon dispersion modeled by the optimized Tersoff potential}

\begin{figure}[h!]
	\centering
	\includegraphics[width=1.0\linewidth]{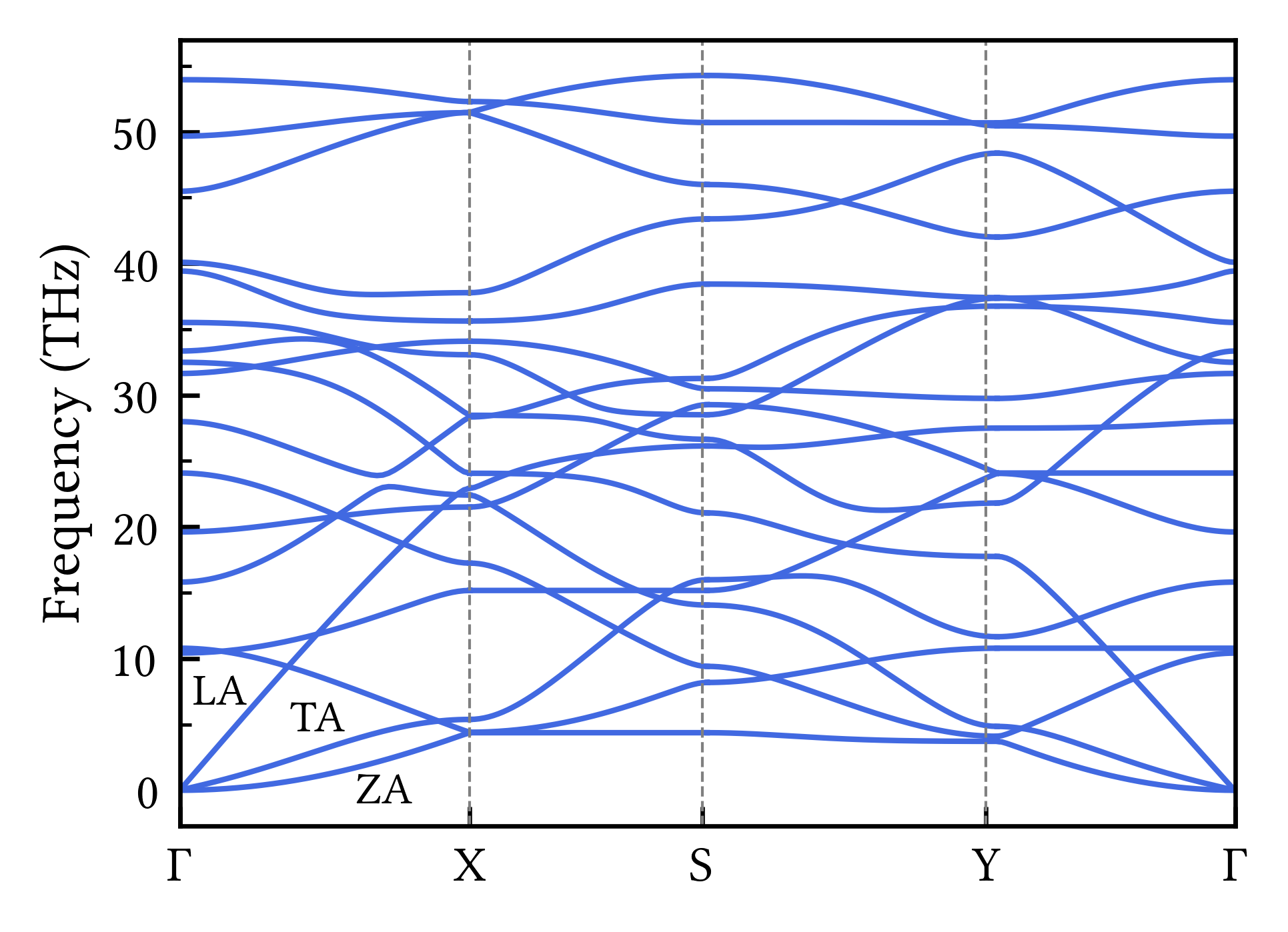}
	\caption{Phonon dispersion for BPN modeled by the optimized Tersoff potential. The absence of negative (imaginary) frequencies indicates the stability of the structure with the chosen potential parameters.}
	\label{fig:phonon-dispersion-sheet}
\end{figure}

In order to verify the accuracy of the optimized Tersoff potential for describing the atomic bonding structure of BPN, we calculated its phonon dispersion. \autoref{fig:phonon-dispersion-sheet} presents the results along high-symmetry points of the Brillouin zone, using the 6-atom unit cell from \autoref{fig:nanotubes}a. The absence of phonon modes with negative (imaginary) frequencies confirms the stability of the BPN crystal structure under the chosen potential. Among the three acoustic phonon modes, the in-plane modes (longitudinal (LA) and transverse (TA)) exhibit linear dispersions, while the out-of-plane mode (flexural (ZA)) shows a quadratic dispersion around the $\Gamma$ point. The optical branches display relatively flat dispersions, indicating small group velocities and minimal contribution to thermal transport. Furthermore, the presence of high-frequency modes ($\sim\SI{50}{\tera\hertz}$), similar to graphene~\cite{Zou2016}, reflects the strong $\mathrm{sp^2}$ bonds between carbon atoms in the BPN structure.

\begin{figure}[h!]
	\centering
	\includegraphics[width=1.0\linewidth]{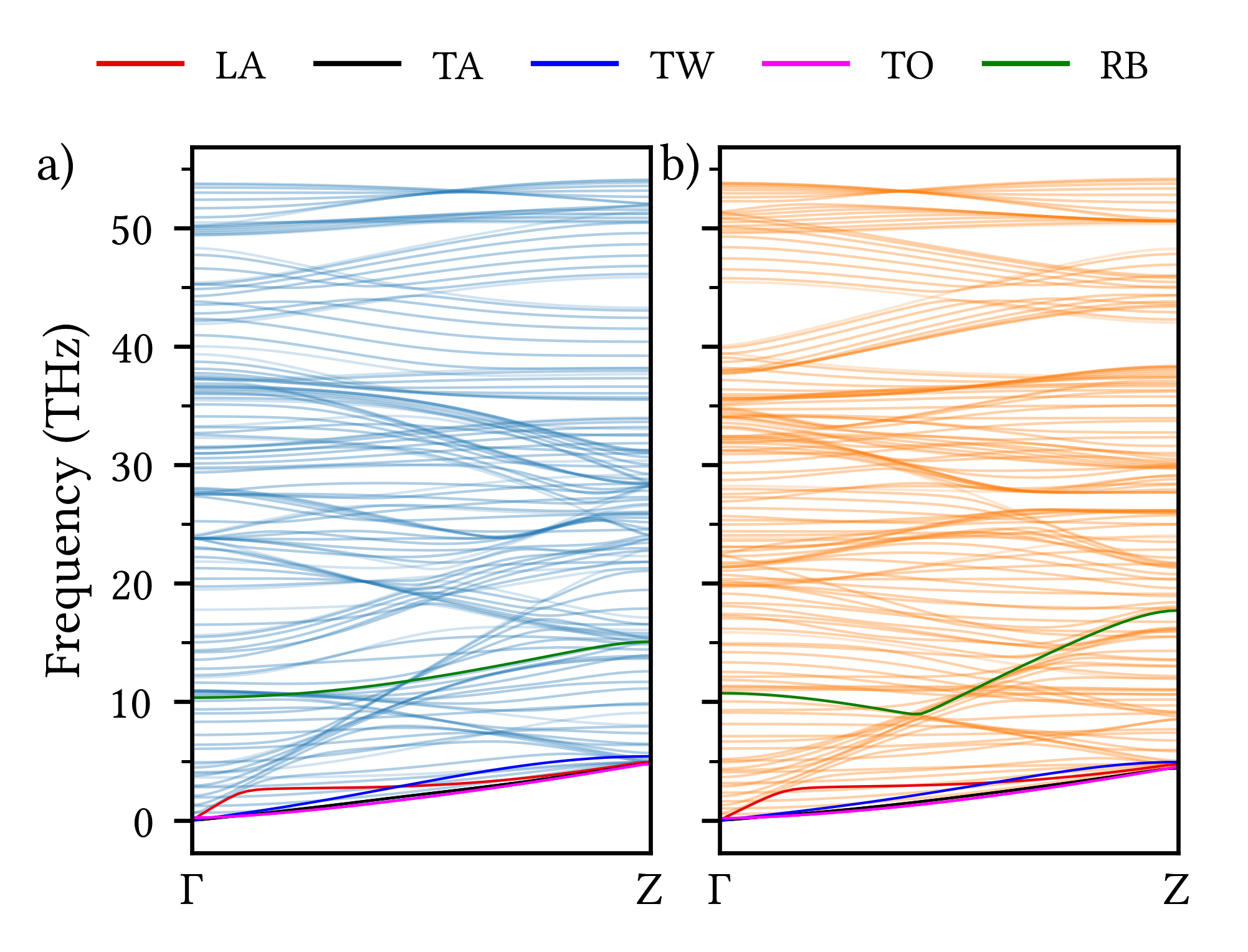}
	\caption{Phonon dispersion for (a) armchair (0,14)- and (b) zigzag (17,0)-BPNNTs modeled by the optimized Tersoff potential. The absence of negative (imaginary) frequencies indicates the stability of the structures with the chosen potential parameters.}
	\label{fig:band_compare}
\end{figure}

Although the phonon dispersion of the planar BPN already provides evidence of structural stability and accurate description of atomic bonding modeled by the optimized Tersoff potential, additional validation through phonon dispersion in BPNNTs is shown in \autoref{fig:band_compare}. The phonon dispersion of BPNNTs exhibits similar characteristics to the planar sheet, with the absence of imaginary frequencies confirming structural stability in the rolled configuration as well. Additionally, due to the circular symmetry of the BPNNTs, the TA mode remains doubly degenerate. Consequently, four acoustic modes are present in BPNNTs, namely, two TA modes, one twisting (TW) mode, and one LA mode. Among the optical phonon branches, the transverse optical (TO) mode and the radial breathing (RB) mode emerge near the $\Gamma$ point at approximately \SI{0.2}{\tera\hertz} and \SI{10}{\tera\hertz}, respectively, consistent with observations in SWCNTs and graphyne nanotubes (GNTs)~\cite{Chen2017}. Furthermore, the preservation of high-frequency optical modes ($\sim\SI{50}{\tera\hertz}$) in the BPNNTs dispersion further demonstrates that the optimized Tersoff potential accurately captures the intrinsic bond stiffness characteristic of the these structures. As a result, the agreement between phonon dispersions of both planar and tubular configurations underscores the versatility and reliability of the potential in describing BPN-based nanostructures.

\subsection{Length Effect on Thermal Conductivity}

The length dependence of thermal conductivity in low-dimensional systems plays a fundamental role in understanding their phonon transport mechanisms. In finite-size systems, phonon confinement within the spatial region between heat reservoirs leads to strong size effects, altering the observed thermal conductivity. This dependence can be described by the following relation~\cite{Schelling2002c}: 
\begin{equation}
\frac{1}{\kappa(L)} = \frac{1}{\kappa_{\infty}} \left( 1 + \frac{\Lambda}{L} \right),
\label{eq:kxlength}
\end{equation}
where $\kappa_{\infty}$ denotes the intrinsic, length-independent thermal conductivity of the material, and $\Lambda$ represents the effective mean free path of heat carriers. By fitting this model to simulation data obtained from systems of varying lengths, both $\kappa_{\infty}$ and $\Lambda$ can be estimated.

\begin{figure}[h!]
	\centering
	\includegraphics[width=1.0\linewidth]{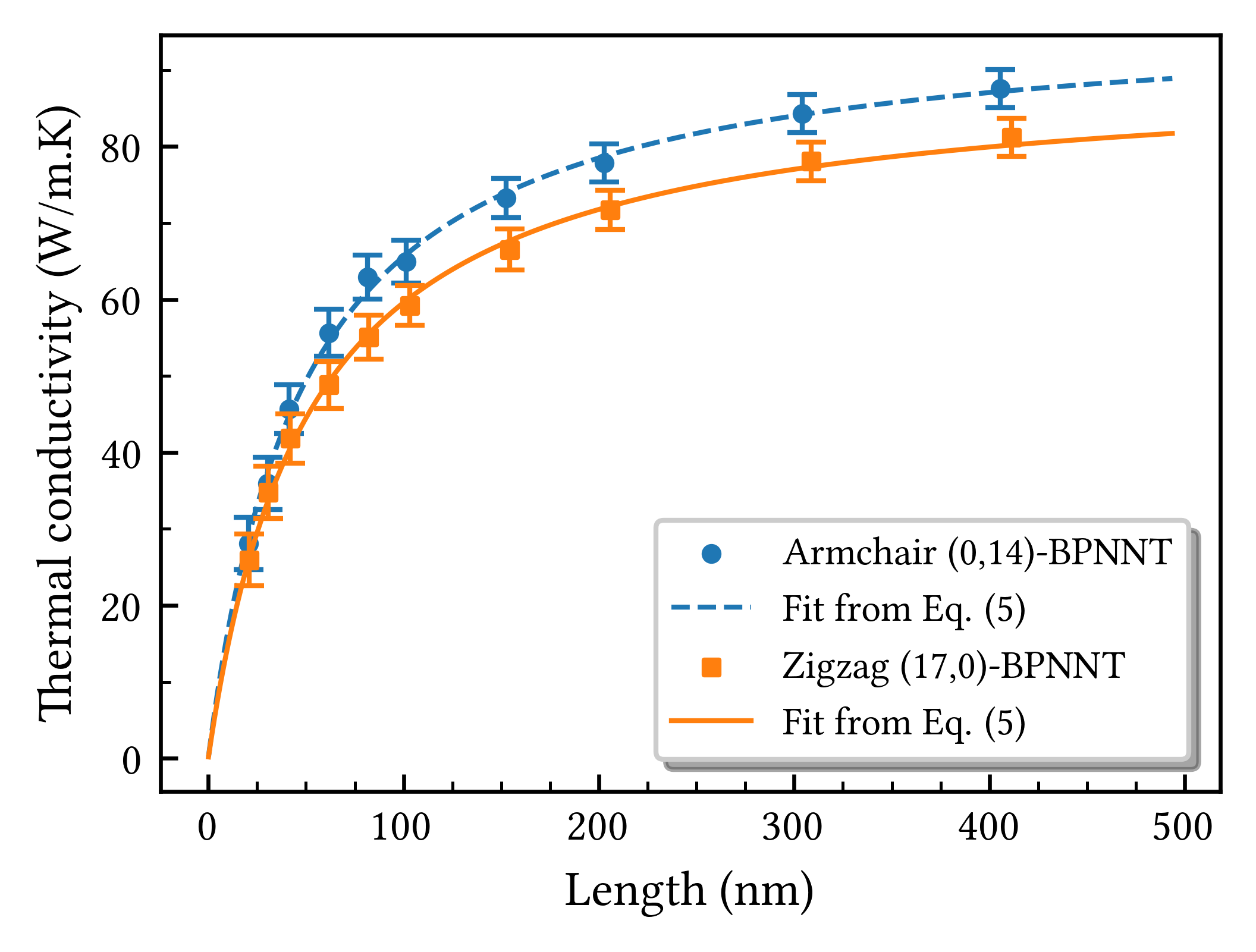}
	\caption{Thermal conductivity as a function of tube length for armchair (0,14)- and zigzag (17,0)-BPNNTs. For both configurations, tubes with diameters of approximately \SI{2}{\nano\meter} are considered.}
	\label{fig:kxlength}
\end{figure}

\autoref{fig:kxlength} illustrates the lattice thermal conductivity of both armchair and zigzag BPNNTs as a function of tube length. The uncertainty in the data points was calculated using uncertainty propagation based on the uncertainties in average heat flux and temperature gradient. At room temperature, the intrinsic lattice thermal conductivity of armchair and zigzag BPNNTs is approximately \SI{98}{\watt\per\meter\per\kelvin} and \SI{90}{\watt\per\meter\per\kelvin}, respectively, with corresponding effective phonon mean free paths of \SI{52}{\nano\meter} and \SI{51}{\nano\meter}. 

\begin{figure}[h!]
	\centering
	\includegraphics[width=1.0\linewidth]{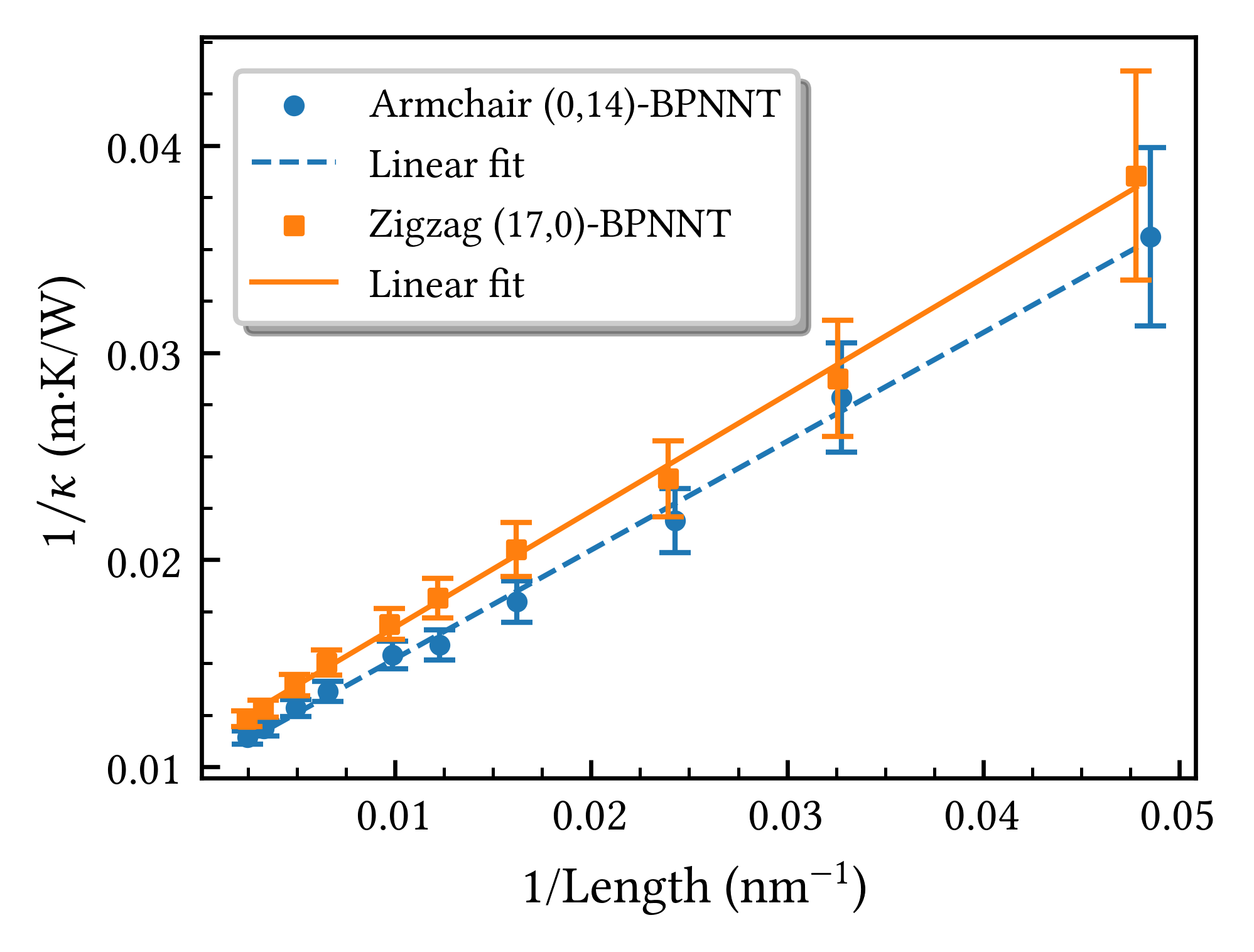}
	\caption{Inverse thermal conductivity as a function of the inverse tube length for armchair (0,14)- and zigzag (17,0)-BPNNTs. For both configurations, tubes with diameters of approximately \SI{2}{\nano\meter} are considered.}
	\label{fig:kxlengthinverse}
\end{figure}

Another commonly used approach in the literature for extracting intrinsic thermal conductivity involves performing a linear regression of $\kappa^{-1}$ versus $L^{-1}$, as illustrated in \autoref{fig:kxlengthinverse}. The linear fit yields thermal conductivity values of \SI{100}{\watt\per\meter\per\kelvin} and \SI{90}{\watt\per\meter\per\kelvin}, along with effective phonon mean free paths of \SI{53}{\nano\meter} and \SI{50}{\nano\meter} for armchair and zigzag BPNNTs, respectively. These results are in good agreement with those obtained from the nonlinear fit, with discrepancies on the order of $2\%$, highlighting the equivalence of both fitting approaches.

\begin{figure}[h!]
	\centering
	\includegraphics[width=1.0\linewidth]{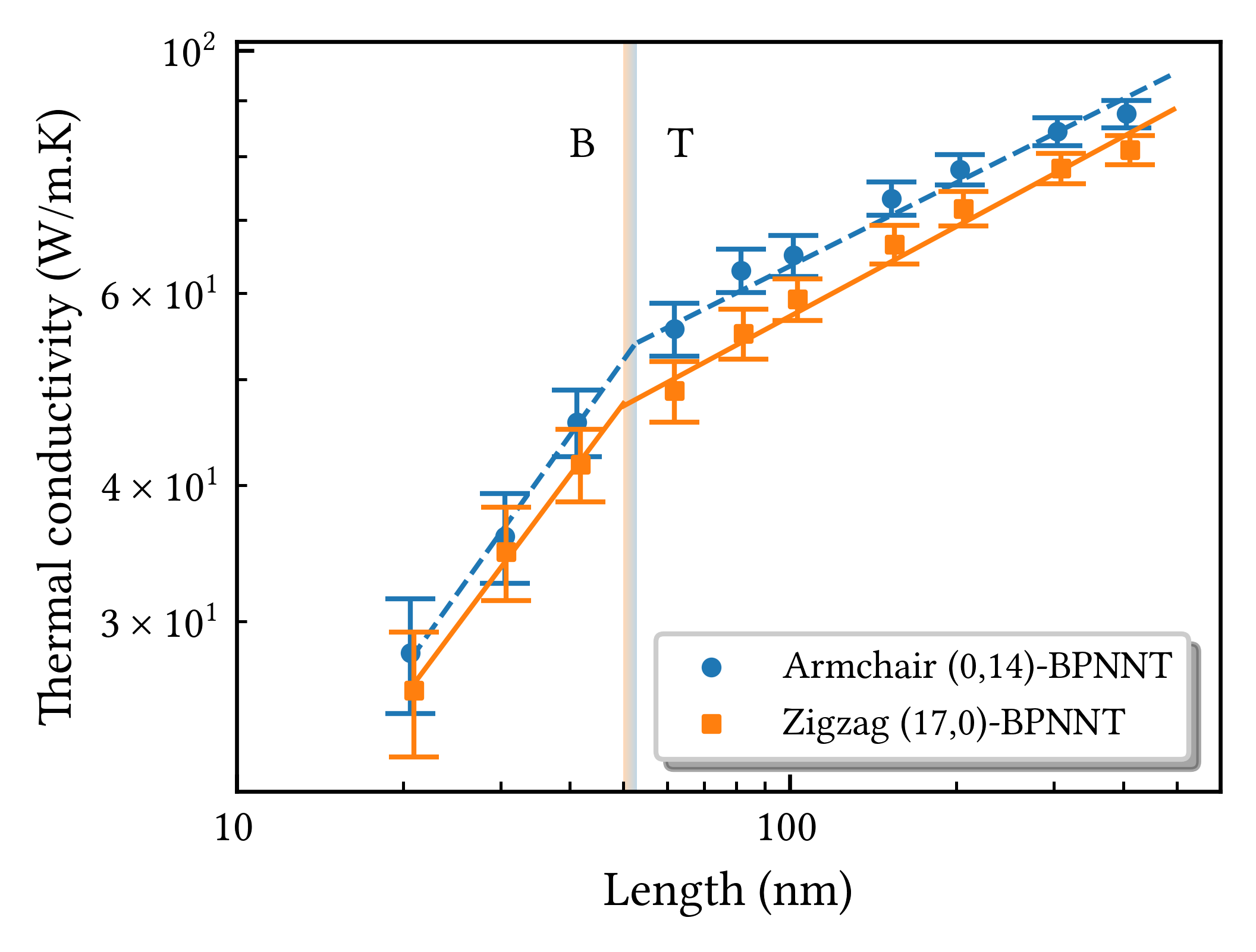}
	\caption{Thermal conductivity as a function of tube length for armchair (0,14)- and zigzag (17,0)-BPNNTs on a log-log scale. For both configurations, tubes with diameters of approximately \SI{2}{\nano\meter} are considered.}
	\label{fig:kxlengthlog}
\end{figure}

\autoref{fig:kxlengthlog} illustrates the lattice thermal conductivity of both armchair and zigzag BPNNTs as a function of tube length on a log-log scale. The behavior of $\kappa(L)$ in \autoref{fig:kxlengthlog} reveals two heat transport regimes in BPNNTs, depending on the tube length. For short tubes $(L <\Lambda)$, thermal transport is primarily governed by phonon-boundary scattering, with phonon-phonon interactions playing a negligible role. In this regime (region B), the phonon mean free path exceeds the system length, leading to predominantly ballistic transport, where thermal conductivity scales with the tube length as $\kappa(L) \propto L^\alpha$, with $\alpha\sim 0.7$. As the tube length increases, the influence of phonon-boundary scattering diminishes, and phonon-phonon scattering gradually becomes the dominant source of thermal resistance. In this ballistic–diffusive transition regime (region T), where the system length is comparable to the phonon mean free path $(L \sim \Lambda)$, heat transport is both ballistic and diffusive, and the thermal conductivity follows a power-law dependence on tube length, $\kappa(L) \propto L^\alpha$, with $\alpha\sim 0.3$. For sufficiently long tubes ($L \gg \Lambda$), phonon-boundary scattering is expected to become negligible, resulting in thermal transport predominantly governed by phonon-phonon interactions and, consequently, independent of tube length. However, due to computational limitations restricting our simulations to tube lengths of only a few nanometers, we were unable to observe this transition to fully diffusive transport in BPNNTs.

The previous results indicate that the thermal conductivity of armchair BPNNTs is slightly higher than that of their zigzag counterparts. Moreover, both values are at least one order of magnitude lower than those reported for SWCNTs employing the same interatomic potential ($\sim\SI{e3}{\watt\per\meter\per\kelvin}$)~\cite{Cao2012}. The moderate thermal conductivity of BPNNTs may facilitate more uniform heat dissipation compared to SWCNTs or graphene, thereby helping to mitigate localized overheating or the formation of thermal ``hot spots''. To gain a better understanding of the mechanisms underlying these differences, it is necessary to analyze their phonon properties.

According to the classical phonon-gas model, the lattice thermal conductivity can be expressed as:
\begin{equation}
    \kappa = \sum_{\mathbf{q}} C_V(\nu) v_g^2(\mathbf{q}, \nu) \tau(\mathbf{q}, \nu),
    \label{eq:kappa_gas}
\end{equation}
where $C_V$ is the volumetric heat capacity, $v_g$ is the phonon group velocity for a phonon with frequency $\nu$ and wave vector $\mathbf{q}$, and $\tau$ is the phonon lifetime. As described by \autoref{eq:kappa_gas}, the phonon group velocity plays a crucial role in determining the lattice thermal conductivity. The phonon group velocity can be derived from the phonon dispersion relation as:
\begin{equation}
    v_g(\nu)=\dfrac{d\nu}{dk},
    \label{eq:group_velocity}
\end{equation}
where $\nu$ represents the frequency of a given mode and $k$ denotes the wave vector.

\begin{figure}[h!]
	\centering
	\includegraphics[width=1.0\linewidth]{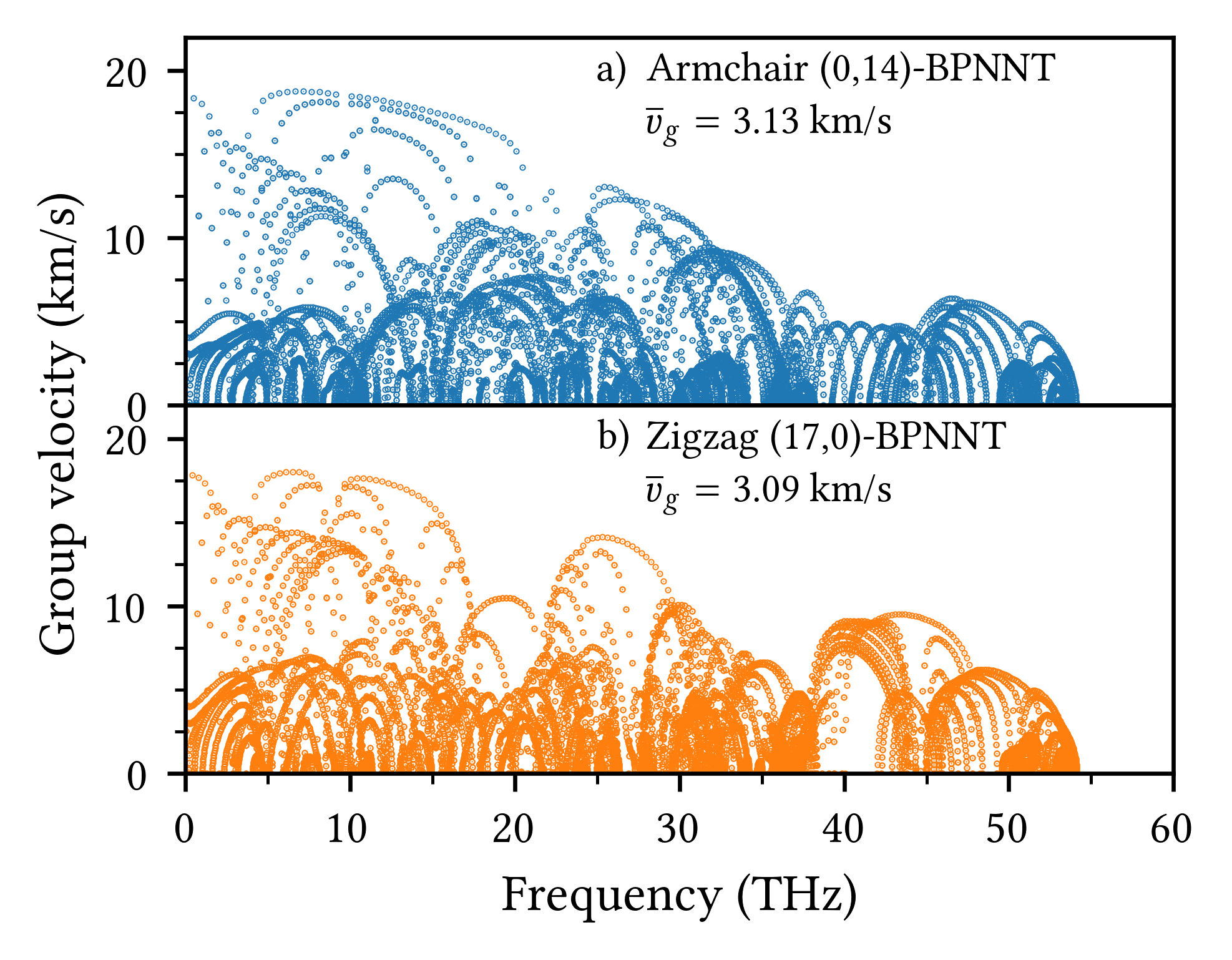}
	\caption{Phonon group velocity for (a) armchair (0,14)- and (b) zigzag (17,0)-BPNNTs. For both configurations, tubes with diameters of approximately \SI{2}{\nano\meter} are considered. The average phonon group velocity for armchair and zigzag BPNNTs is \SI{3.13}{\kilo\meter\per\second} and \SI{3.09}{\kilo\meter\per\second}, respectively.}
	\label{fig:grou_velocity_compare}
\end{figure}

\autoref{fig:grou_velocity_compare} compares the phonon group velocities of both armchair and zigzag BPNNTs. The calculated average group velocities are \SI{3.13}{\kilo\meter\per\second} and \SI{3.09}{\kilo\meter\per\second} for armchair and zigzag BPNNTs, respectively. These results suggest that the slightly higher thermal conductivity observed in armchair BPNNTs, compared to their zigzag counterparts, can be attributed to their higher average phonon group velocity.

\begin{figure}[h!]
	\centering
	\includegraphics[width=1.0\linewidth]{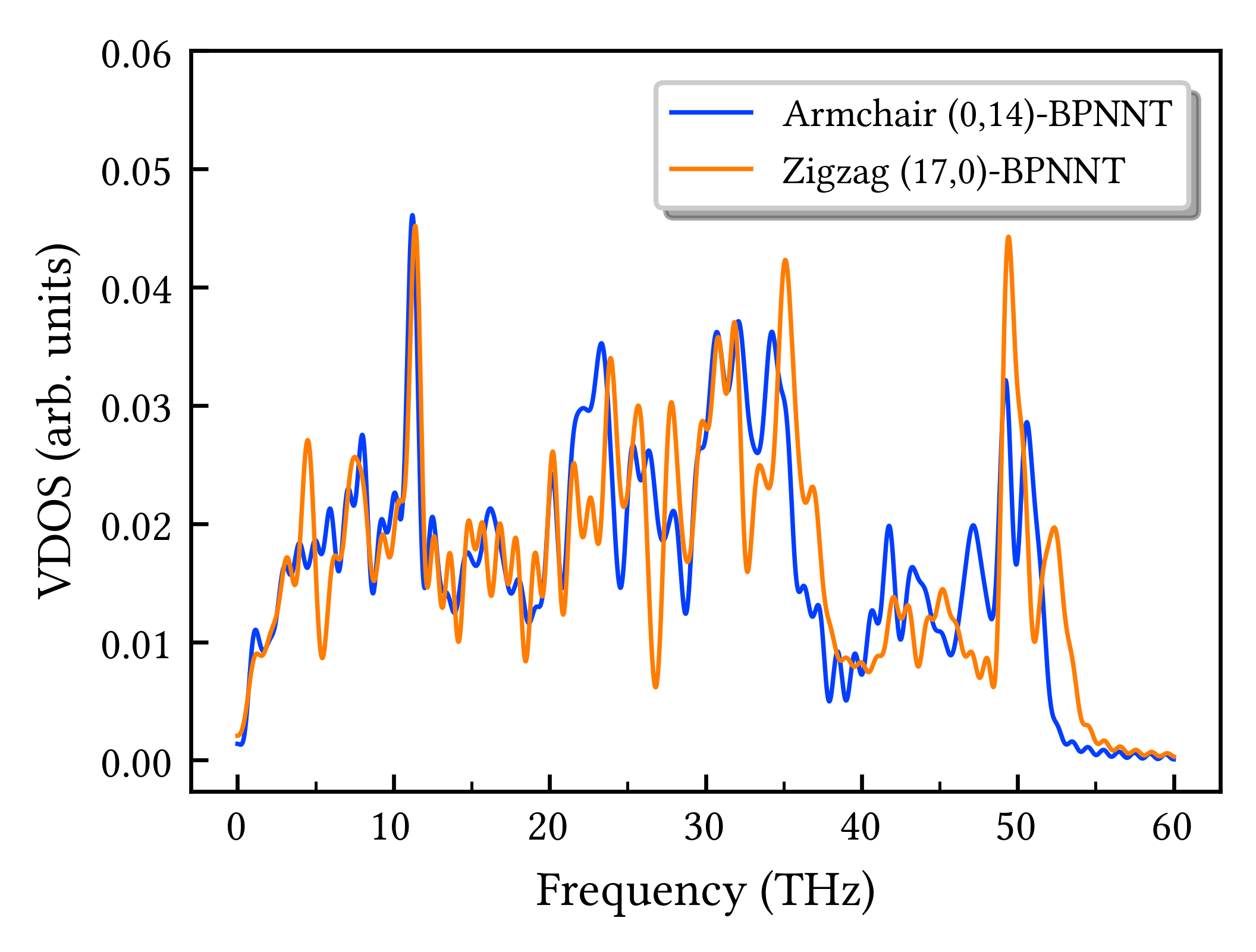}
	\caption{VDOS of armchair (0,14)- and zigzag (17,0)-BPNNTs calculated using \autoref{eq:vdos}. For both configurations, tubes with diameters of approximately \SI{2}{\nano\meter} and lengths of \SI{100}{\nano\meter} are considered.}
	\label{fig:vdosxquirality}
\end{figure}

To further elucidate the influence of chirality on the thermal conductivity of BPNNTs, their VDOS is depicted in \autoref{fig:vdosxquirality}. The overall spectral shape is quite similar for both configurations. However, the armchair BPNNT exhibit more prominent peaks in the VDOS for low-frequency modes, along with a noticeable shift in the high-frequency modes. As shown in \autoref{fig:grou_velocity_compare}, high-frequency modes are associated with lower group velocities, whereas low-frequency modes exhibit higher group velocities and, thus, contribute more significantly to thermal transport. Therefore, the slightly higher thermal conductivity observed in armchair BPNNTs compared to zigzag ones can be attributed to a combination of a greater number of low-frequency modes and their comparatively higher average group velocity.

Finally, the substantial difference between the thermal conductivity of BPNNTs and SWCTNs arises primarily from the distinct group velocities observed between these structures. In SWCNTs, the average group velocity is on the order of $\SI{e{1}}{\kilo\meter\per\second}$~\cite{Bing2009theoretical}, significantly higher than that of BPNNTs. Consequently, the lower group velocities of BPNNTs directly result in their reduced lattice thermal conductivity compared to SWCNTs.

\subsection{Diameter Effect on Thermal Conductivity}

\begin{figure}[h!]
	\centering
	\includegraphics[width=1.0\linewidth]{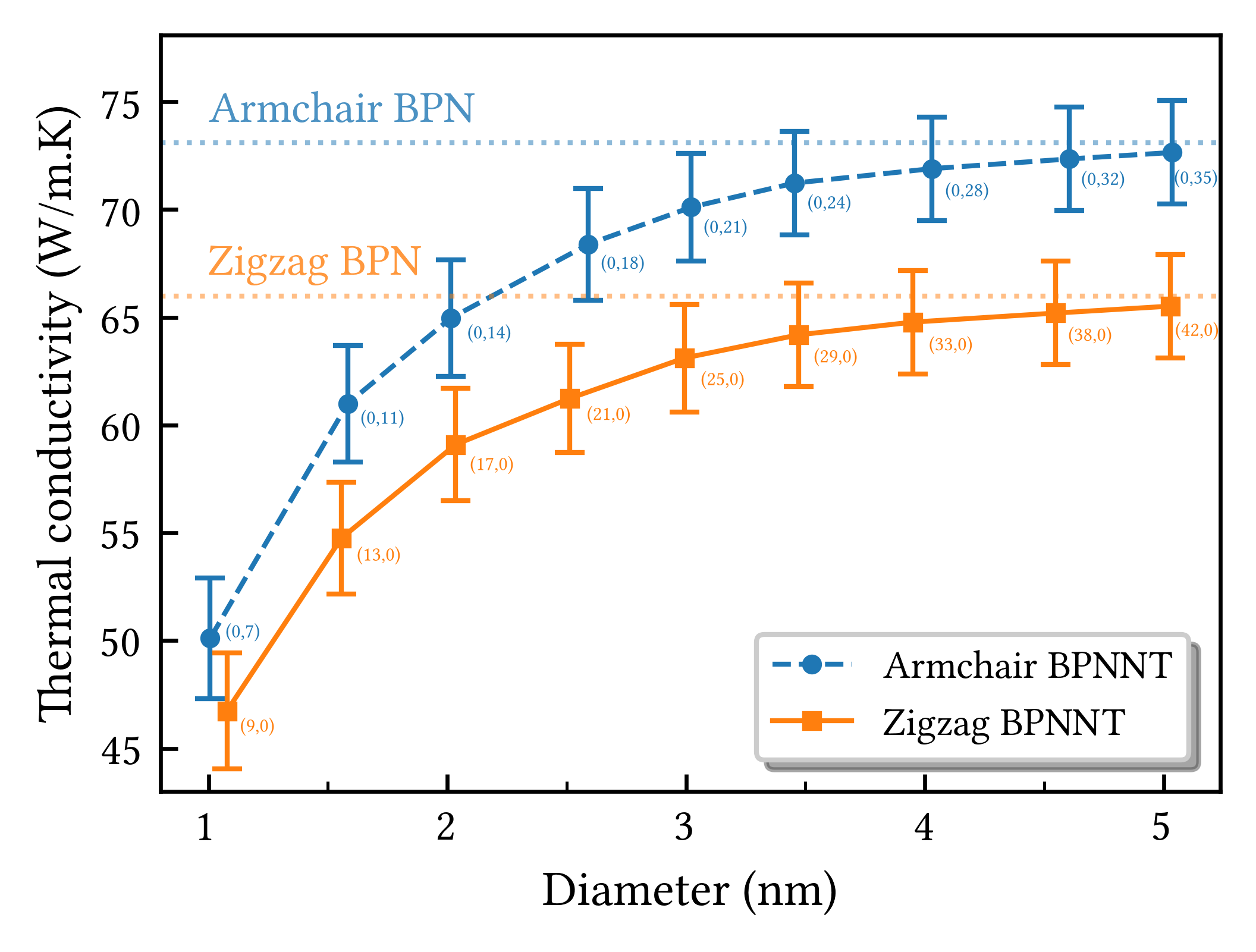}
	\caption{Thermal conductivity as a function of tube diameter for armchair and zigzag BPNNTs. All tubes considered have lengths of approximately \SI{100}{\nano\meter}. The corresponding chiral indices for each tube are indicated. The horizontal dotted lines indicate the thermal conductivity of BPN nanoribbons with comparable length and cross-sectional area to that of BPNNTs with a diameter of \SI{5}{\nano\meter}.}
	\label{fig:kxdiameter}
\end{figure}

\autoref{fig:kxdiameter} illustrates the dependence of the thermal conductivity on the diameter for both armchair and zigzag BPNNTs. In addition, thermal conductivity calculations were performed for BPN monolayers with armchair and zigzag edges, using a methodology analogous to that employed for the tubes. The BPN monolayers, with a length of \SI{100}{\nano\meter} and a cross-sectional area similar to that of BPNNTs with a diameter of \SI{5}{\nano\meter}, were simulated under periodic boundary conditions applied solely along the length direction. This configuration permits a direct comparison of the curvature effects on thermal transport.

A detailed analysis of \autoref{fig:kxdiameter} reveals that the thermal conductivity increases nearly exponentially with increasing tube diameter for both types of BPNNTs. For example, when the diameter increases from approximately \SI{1}{\nano\meter} to \SI{5}{\nano\meter}, the thermal conductivity of armchair BPNNTs rises by about $45\%$, while that of zigzag BPNNTs increases by approximately $40\%$. At smaller diameters (up to \SI{3}{\nano\meter}), the thermal conductivity exhibits a strong diameter dependence, increasing significantly with larger diameters. However, as the tube diameter increases further, the thermal conductivity approaches a plateau, becoming nearly independent of diameter and converging towards the value characteristic of the corresponding planar structure. This behavior can be attributed primarily to the reduction in the surface area-to-volume ratio as the tube diameter increases, which leads to a decrease in phonon-boundary scattering. As a result, phonons can travel longer distances without scattering, thereby increasing their mean free path and enhancing the thermal conductivity, as described by \autoref{eq:kappa_gas}. Similar trends have been reported for SWCNTs, supported by molecular dynamics simulations employing the optimized Tersoff potential~\cite{Cao2012}, density functional theory (DFT) calculations~\cite{Yue2015}, and experimental measurements~\cite{Zhao2025}.

\begin{figure}[h!]
	\centering
	\includegraphics[width=1.0\linewidth]{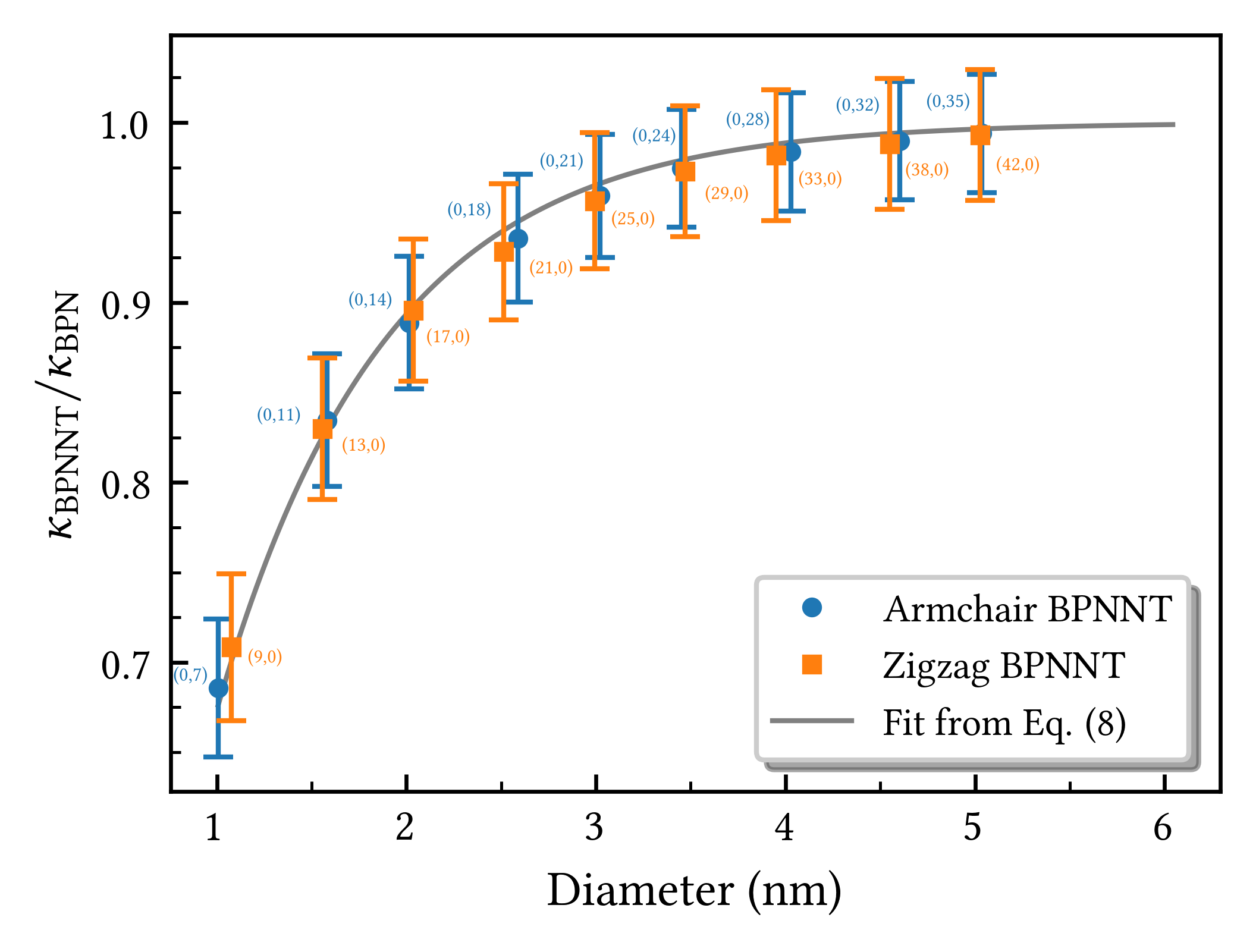}
	\caption{Diameter dependence of the thermal conductivity for armchair and zigzag BPNNTs with lengths of approximately \SI{100}{\nano\meter}. All values are normalized by the thermal conductivity of the corresponding BPN nanoribbon of the same length, denoted as $\kappa_\mathrm{BPN}$. The corresponding chiral indices for each tube are indicated.}
	\label{fig:kxdiameter-fit}
\end{figure}

To further investigate the diameter dependence of thermal conductivity, \autoref{fig:kxdiameter-fit} presents the thermal conductivity of armchair and zigzag BPNNTs normalized by the thermal conductivity of the corresponding BPN monolayer of the same length ($\kappa_\mathrm{BPN}$). Based on the preceding discussion, the thermal conductivity of BPNNTs is expected to vary with diameter according to::
\begin{equation}
    \kappa_\mathrm{BPNNT}(d) \approx\kappa_\mathrm{BPN}(1-e^{-\gamma d}),
    \label{eq:kappaxd}
\end{equation}
where $\gamma$ is a fitting parameter with units of inverse length. By fitting the data presented in \autoref{fig:kxdiameter-fit} to \autoref{eq:kappaxd}, we obtain $\gamma=\SI{1.12}{\per\nano\meter}$ for both types of BPNNTs. In Ref.~\cite{Cao2012}, the authors reported $\gamma=\SI{0.75}{\per\nano\meter}$ for SWCNTs. This comparison indicates that the thermal conductivity of BPNNTs increases more rapidly with diameter than that of SWCNTs.

\begin{figure}[h!]
	\centering
	\includegraphics[width=1.0\linewidth]{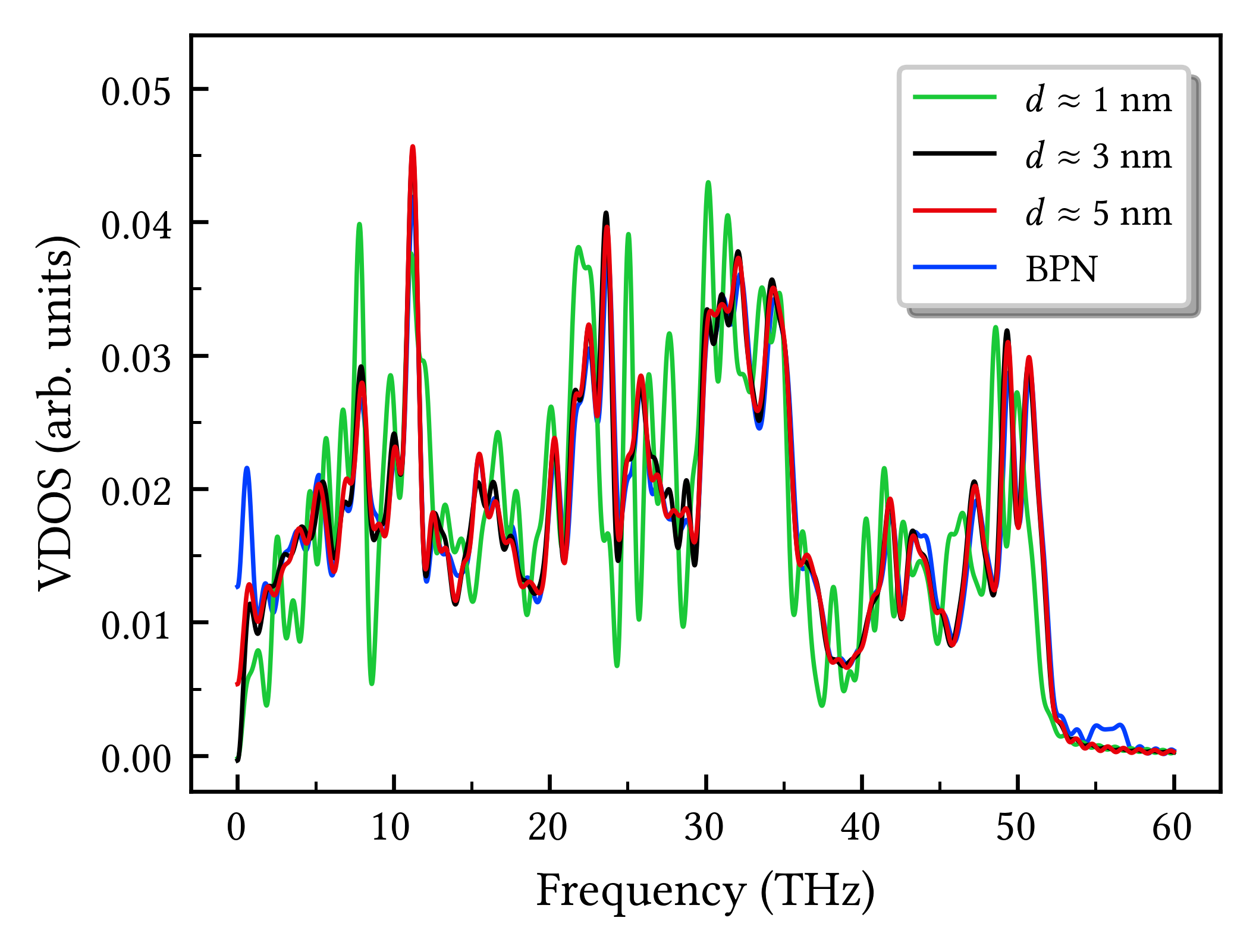}
	\caption{VDOS of armchair BPNNTs with different diameters (all with a length of \SI{100}{\nano\meter}), and a BPN nanoribbon with a width chosen to match the cross-sectional area of the widest tube. The VDOS is normalized on a per-atom basis, ensuring comparability across different tube diameters.}
	\label{fig:vdosxdiameter}
\end{figure}

\begin{figure}[h!]
	\centering
	\includegraphics[width=1.0\linewidth]{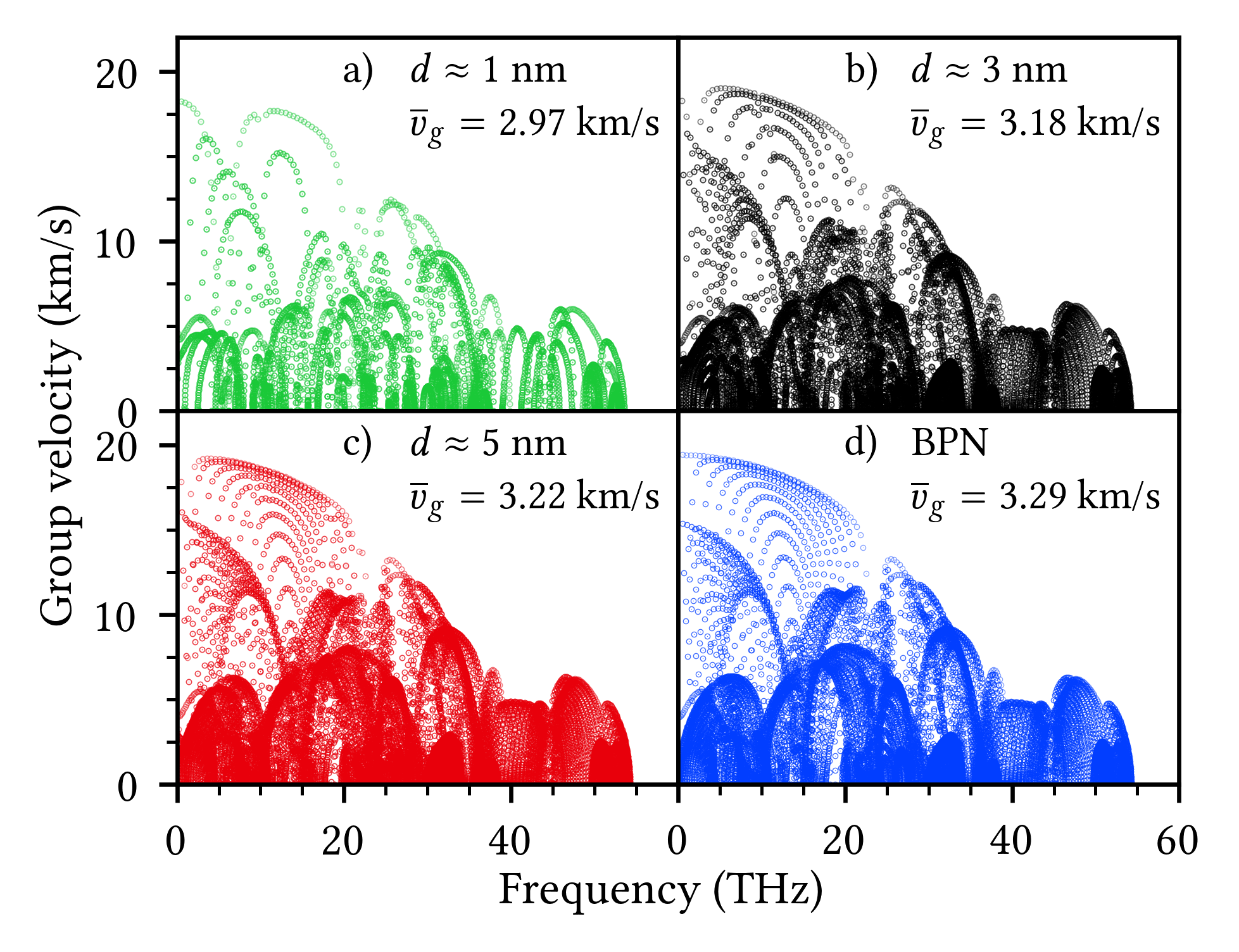}
	\caption{Phonon group velocity of armchair BPNNTs with diameters of approximately (a) \SI{1}{\nano\meter}, (b) \SI{3}{\nano\meter}, and (c) \SI{5}{\nano\meter} (all with a length of \SI{100}{\nano\meter}), and (d) a BPN nanoribbon with a width selected to match the cross-sectional area of the widest nanotube.}
	\label{fig:groupvelocityxdiameter}
\end{figure}

To gain further insight into the diameter dependence of thermal conductivity, the VDOS of armchair BPNNTs with different diameters is shown in \autoref{fig:vdosxdiameter} as a representative case. Significant differences are observed between the VDOS of tubes with diameters of \SI{1}{\nano\meter} and \SI{3}{\nano\meter}, whereas the VDOS of the \SI{5}{\nano\meter} BPNNT closely resembles that of the \SI{3}{\nano\meter} one. Moreover, the VDOS of the \SI{5}{\nano\meter} tube becomes nearly indistinguishable from that of a BPN nanoribbon, indicating a convergence toward planar behavior. This behavior is consistent with the thermal conductivity trend shown in \autoref{fig:kxdiameter}, where $\kappa$ increases sharply as the diameter increases from \SI{1}{\nano\meter} to \SI{3}{\nano\meter}, and then gradually saturates for larger diameters. Further analysis of \autoref{fig:vdosxdiameter} reveals that tubes with larger diameters exhibit a higher density of vibrational modes in the low-frequency region (below \SI{5}{\tera\hertz}). Additionally, \autoref{fig:groupvelocityxdiameter} indicates that the average phonon group velocity increases with diameter, which in turn contributes to the enhancement of thermal conductivity. For instance, the average group velocity increases by approximately $7\%$ as the diameter increases from \SI{1}{\nano\meter} to \SI{3}{\nano\meter}, but only by $1\%$ from \SI{3}{\nano\meter} to \SI{5}{\nano\meter}. Notably, the group velocity profile of the \SI{5}{\nano\meter} tube also aligns with that of the nanoribbon, reinforcing the interpretation that BPNNTs with sufficiently large diameter approach the thermal transport characteristics of their planar 1D counterpart.

\subsection{Temperature Effect on Thermal Conductivity}
\begin{figure}[h!]
	\centering
	\includegraphics[width=1.0\linewidth]{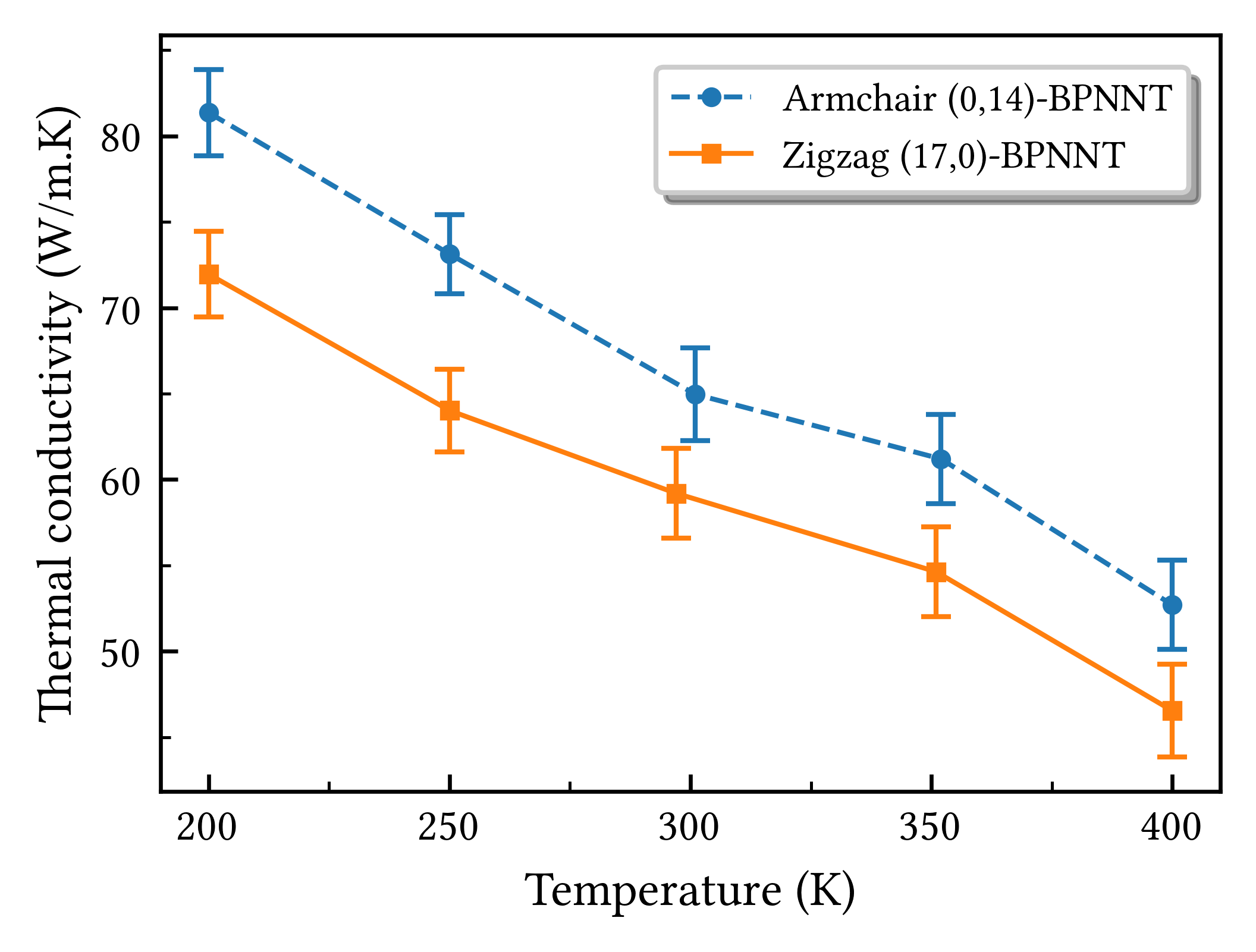}
	\caption{Thermal conductivity as a function of temperature for armchair (0,14)- and zigzag (17,0)-BPNNTs. For both configurations, tubes with diameters of approximately \SI{2}{\nano\meter} and lengths of \SI{100}{\nano\meter} are considered.}
	\label{fig:kxtemperature}
\end{figure}

\autoref{fig:kxtemperature} presents the thermal conductivity of armchair and zigzag BPNTs at different temperatures. The results indicate a similar monotonic decrease in thermal conductivity for both configurations with increasing temperature. For example, increasing the temperature from \SI{200}{\kelvin} to \SI{400}{\kelvin} results in a reduction of approximately $35\%$ in the thermal conductivity of both armchair and zigzag BPNTs. This behavior has also been observed in BPN monolayers~\cite{HamedMashhadzadeh2022, Yang2023, Veeravenkata2021}, graphene~\cite{Chen2012}, SWCNTs~\cite{Marconnet2013}, and carbon nanoscrolls (CNSs)~\cite{deLima2025}.

\begin{figure}[h!]
	\centering
	\includegraphics[width=1.0\linewidth]{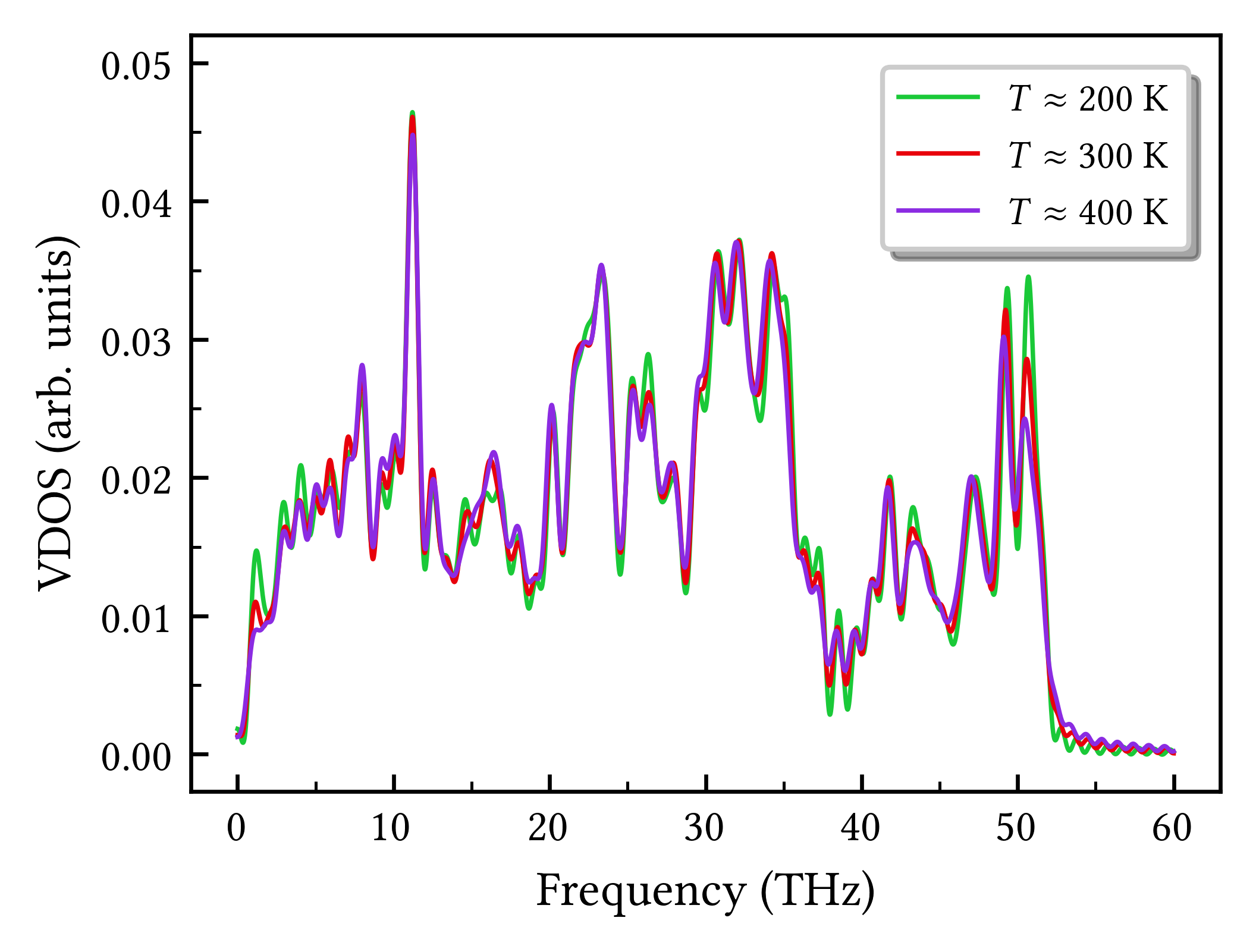}
	\caption{VDOS of armchair (0,14)-BPNNTs at different temperatures. For all configurations, tubes with diameters of approximately \SI{2}{\nano\meter} and lengths of \SI{100}{\nano\meter} are considered.}
	\label{fig:vdosxtemperature}
\end{figure}

To gain further insights into the temperature dependence of thermal conductivity, the VDOS of armchair BPNNTs at different temperatures is presented in \autoref{fig:vdosxtemperature} as a representative case. While the overall spectral shape remains consistent, the VDOS at lower temperatures exhibit more pronounced peaks, especially in the low-frequency (below \SI{5}{\tera\hertz}) and high-frequency ($\sim\SI{50}{\tera\hertz}$) regions. This behavior indicates a suppression of both low- and high-frequency vibrational modes with increasing temperature. However, high-frequency modes contribute minimally to the thermal conductivity due to their short-ranged, localized nature and low group velocity. In contrast, low-frequency modes are long-ranged, delocalized, and possess higher group velocities, making them the dominant heat carriers. Consequently, the attenuation of these low-frequency modes with increasing temperature results in reduced thermal conductivity. Additionally, at higher temperatures, the intensification of phonon-phonon interactions leads to enhanced phonon scattering. This increased scattering reduces the mean free path of phonons, thereby decreasing thermal conductivity~\cite{Veeravenkata2021}.

\section{Summary and conclusions}

In this work, non-equilibrium molecular dynamics simulations were performed to investigate the thermal transport properties of nanometer-long BPNNTs. Particular emphasis was placed on understanding the influence of tube length, diameter, and temperature on the lattice thermal conductivity. At room temperature, the intrinsic lattice thermal conductivity of armchair and zigzag BPNNTs is approximately \SI{100}{\watt\per\meter\per\kelvin} and \SI{90}{\watt\per\meter\per\kelvin}, respectively. These values are at least one order of magnitude lower than those reported for SWCNTs using the same interatomic potential. This significant difference is primarily attributed to the lower phonon group velocities in BPNNTs compared to SWCNTs.

The length dependence of thermal conductivity reveals that the transition from predominantly ballistic to ballistic-diffusive transport occurs at lengths of approximately \SI{53}{\nano\meter} and \SI{50}{\nano\meter} for armchair and zigzag BPNNTs, respectively. These transition lengths are considerably shorter than those reported for SWCNTs, further supporting the interpretation that differences in thermal conductivity between BPNNTs and SWCNTs arise from their distinct phonon properties.

Regarding the effect of tube diameter, our results show that the thermal conductivity of both armchair and zigzag BPNNTs increases with increasing diameter, asymptotically approaching the thermal conductivity of a BPN nanoribbon. This behavior is attributed to the gradual increase in phonon group velocity as the diameter increases, reflecting a convergence toward the planar geometry.

Finally, the thermal conductivity of both BPNNT types is found to decrease with increasing temperature. This reduction is attributed to the suppression of low-frequency vibrational modes and the enhancement of phonon-phonon scattering processes at higher temperatures, both of which reduce the phonon mean free path and, consequently, the thermal conductivity.

In summary, this study provides deeper insight into the thermal transport properties of BPNNTs and expands the understanding of thermal properties within the carbon-based nanotube family, with implications for thermal management and energy-related applications.

\begin{acknowledgments}
We thank the Coaraci Supercomputer for computer time (Fapesp grant \#2019/17874-0) and the Center for Computing in Engineering and Sciences at Unicamp (Fapesp grant \#2013/08293-7).
\end{acknowledgments}



\bibliography{manuscript.bib}
\end{document}